\documentclass[prb,showpacs,twocolumn,superscriptaddress,floatfix,longbibliography,nofootinbib]{revtex4-2}
\usepackage{graphicx,amsfonts,amssymb,amsmath,xspace,dsfont,bm,bbm}
\usepackage[colorlinks=true,citecolor=green,linkcolor=red,urlcolor=blue]{hyperref}
\usepackage{subfigure}
\usepackage{blindtext}
\usepackage{tabularx}
\usepackage{circuitikz}

\usepackage{color}
\definecolor{g}{rgb}{.1,0.4,.1} % {.0,0.7,.5}
\definecolor{b}{rgb}{0,0.2,1}
\definecolor{rouge}{rgb}{0.82,0.,0.}
\definecolor{vert}{rgb}{0.,0.82,0.}
\definecolor{orange}{rgb}{1,0.5,0.}
\definecolor{bleu}{rgb}{0.,0.,0.82}
\definecolor{m}{rgb}{0.82,0.,0.82}
\definecolor{vert2}{rgb}{0.,0.5,0.}
\definecolor{rougeclair}{rgb}{1.0,0.7,0.7}

\newcolumntype{Y}{>{\centering\arraybackslash}X}

\begin{document}

\title{Disorder-driven stochastic dynamics in Mott resistive-switching systems}

\author{David J. Alspaugh}
\affiliation{Physics Department, University of California, San Diego, 9500 Gilman Dr., La Jolla, CA 92093, USA}

\author{Lorenzo Fratino}
\affiliation{Laboratoire de Physique Th{\'e}orique et Mod{\'e}lisation, CNRS, CY Cergy Paris Universit{\'e}, 95302 Cergy-Pontoise Cedex, France}

\author{Nareg Ghazikhanian}
\affiliation{Physics Department, University of California, San Diego, 9500 Gilman Dr., La Jolla, CA 92093, USA}

\author{Ivan K. Schuller}
\affiliation{Physics Department, University of California, San Diego, 9500 Gilman Dr., La Jolla, CA 92093, USA}

\author{Marcelo Rozenberg}
\affiliation{Laboratoire de Physique des Solides, CNRS-UMR 8502, Universit{\'e} Paris-Saclay, Orsay 91405, France}

\date{February 14, 2026}

\begin{abstract}

{Controlled disorder in correlated materials provides a new route to emergent stochastic dynamics in neuromorphic hardware. Here we show that focused ion beam irradiation in VO$_{2}$- and V$_{2}$O$_{3}$-based resistive-switching oscillators induces a transition from regular periodic oscillations to strongly irregular stochastic firing, while simultaneously reducing the required switching energy by orders of magnitude. Under an applied electric field, these materials undergo a volatile insulator-to-metal transition characterized by the formation of percolating metallic filaments within an insulating bulk. Using numerical simulations based on the Mott resistor network, we demonstrate that defect-induced modifications to filament nucleation and stability drive these devices into stochastic oscillatory regimes. These results are validated by experimental measurements on irradiated VO$_{2}$ and V$_{2}$O$_{3}$ devices.}

\end{abstract}


\maketitle

\section{Introduction}

The phenomenon of electrically driven resistive switching (RS) in materials that undergo an insulator-to-metal phase transition (IMT) has been the subject of an intense research focus over the past several years~\cite{Mazza16,Stoliar14,Camjayi14,Zimmers13}. This is in part due to the potential application of these materials to act as artificial neurons and synapses for hardware-level implementations of neuromorphic computing~\cite{Stoliar17,delValle18,delValle20,Wu24}. The physical origins of the RS behavior, as well as the materials themselves, are myriad. Such mechanisms driving the IMT include ionic drift~\cite{Waser07}, crystallization and amorphization~\cite{Wong10,Zhang19}, or electrical triggering via voltage~\cite{Duchene71,Ghazikhanian23}. Typically, the RS behavior manifests via the formulation a percolating metallic filamentary domain within the otherwise insulating material~\cite{Kisiel25,Pofelski25}. The growth of these filaments heavily depends on the details of the device morphology in question, as well as the switching power necessary for filament creation~\cite{Rocco22,Stoliar13,delValle19,Tesler18,Qiu23}. Furthermore, nanoscale features such as grain size, local defects, and interfacial strain can dramatically affect the filament growth behavior and power necessary for RS~\cite{Shabalin20,delValle18,delValle17}.

Among the class of materials described above, the vanadates are particularly interesting. These materials, such as VO$_{2}$ and V$_{2}$O$_{3}$, exhibit ``volatile" RS, where after the resistive collapse, the system is able to spontaneously return to its original insulating state, i.e., the metallic filament reabsorbs. Both VO$_{2}$ and V$_{2}$O$_{3}$ are strongly correlated Mott materials exhibiting a first-order IMT at 150 K and 340 K respectively~\cite{Morin59,Park13,Imada98,Cheng21,Lee17}. When a strong electric field is applied below the bulk IMT temperature, these materials also exhibit electrically stimulated partial metallization in the form of percolating metallic filaments within an otherwise insulating bulk. When the applied voltage is removed, these filamentary domains disappear, giving rise to the RS behavior. Dynamic self-oscillations in VO$_{2}$ and V$_{2}$O$_{3}$ may be realized via the assembly of a Pearson-Anson type circuit~\cite{Adda22,Qiu23,Pearson21}. A Pearson-Anson circuit is a simple RC-based nonlinear oscillator used to probe and exploit RS or memristive behavior. In RS devices built from these materials, either the intrinsic self capacitance of the device (due to its capacitor-like geometry) or an external capacitance can give rise to sustained oscillations as a constant current is applied. The mechanism is as follows: As a constant current or voltage is applied to the device, the capacitor begins to charge. This increases the voltage across both the capacitor and the Mott material in parallel. When the voltage is high enough, the material undergoes a resistive collapse via the formation of a metallic filament bridging the two electrodes of the device. This functionally acts as a short to the capacitor, which immediately discharges through the material resulting in a large current spike. After the discharge, the voltage across the material quickly drops towards zero, leading to the reabsorption of the metallic filament as the insulating phase is recovered. Once again, the capacitor begins to charge, thereby resulting in a sustained cycle of current pulses through the device. The oscillation regime described above occurs only within a certain dc bias range, and outside this range the RS material is either fully insulating or metallic. Inducing these dynamic self-oscillating RS regimes within these materials has several applications in the field of neuromorphic computing, such as allowing one to mimic the spiking behavior of biological neurons~\cite{Gerstner14,Rabinovich06,Yi2018}.

Previous studies have shown that the global properties of RS and filament formation in VO$_{2}$ and V$_{2}$O$_{3}$ may be reliably manipulated via focused ion beam (FIB) irradiation. Through FIB irradiation, not only can the location and shape of the conducting filaments be controlled, but in addition the necessary threshold voltages required to induce filament formation can be substantially reduced~\cite{Lesueur93,Xiang21,Ghazikhanian23,Ghazikhanian25}. These properties are likely modified due to changes in carrier concentration, mobility, and phase change behavior~\cite{Mei22,Kalcheim20}.

The filamentary resistive collapse has been studied numerically in Ref.~\onlinecite{Rocco22} providing many insights. In particular, it has been shown that the percolation of filaments near the threshold voltage required for the IMT is intrinsically stochastic, and qualitatively similar to the stochastic firing of biological neurons. This stochasticity is borne via both electrical and thermal effects in the form of electrical current density focusing and Joule self-heating, and has been shown to closely reproduce the experimental escape rate behavior of VO$_{2}$ and V$_{2}$O$_{3}$-based devices. However, the consequences of disorder within these systems has been much less investigated. Controlled disorder can be implemented by focused ion beam irradiation, and has previously been shown to have a useful impact in the fabrication and engineering of neuromorphic devices. Therefore a theoretical treatment in modeling the effects of disorder in the conductive filament formation in volatile RS materials is important, and may provide the guidance needed in the understanding of these non-linear phenomena.

In this work we numerically and experimentally study the effects of FIB irradiation on the dynamic self oscillations of Mott spiking neuron circuits built with VO$_{2}$ and V$_{2}$O$_{3}$. Chiefly, we demonstrate that FIB irradiation can induce stochastic and non-periodic voltage oscillations within these devices, sharing interesting similarities with certain behaviors of biological neurons~\cite{Bianchi2012}. Here, we take the term “stochasticity” to explicitly refer to the standard deviation of the average inter-spike time period within a spiking neuron device. We numerically simulate the RS materials by extending the Mott resistor network (MRN), which has previously been used to analyze metallic filament formation within VO$_{2}$ and V$_{2}$O$_{3}$~\cite{Rocco22,Stoliar13,delValle19,Tesler18,Qiu23}, to model the effects of ion irradiation. In Section~\ref{sec2} we introduce this model and describe how it may be applied towards the analysis of FIB irradiation. In particular, we demonstrate how localized defects arising from the implanted ions can give rise to modified RS behavior. In Section~\ref{sec3} we analyze the results of this model when applied to spiking neuron circuits and demonstrate that certain densities of defects can sensitively affect the self-oscillating regime, and even introduce large stochasticity for higher applied dc bias voltages. In addition, we validate our model's predictions by comparing them with experimental measurements performed on irradiated devices featuring both VO$_{2}$ and V$_{2}$O$_{3}$. In Section~\ref{sec4} we discuss our results and comment on the potential application of these devices in neuromorphic circuitry. More broadly, our results demonstrate that disorder can act as a constructive control parameter for nonlinear nonequilibrium dynamics in correlated materials, establishing a general mechanism for tuning stochasticity in resistive switching systems. Because the modeling framework developed here is not specific to the vanadates, these ideas extend naturally to a broad class of correlated oxides and volatile RS materials.

\section{Mott resistor network}
\label{sec2}

In this section we introduce the MRN, which is a numerical method that has previously been used to simulate the growth of metallic filaments within RS materials exhibiting both first and second order IMT's~\cite{Rocco22,Stoliar13,delValle19,Tesler18,Qiu23,Adda22}. Here in this work we focus on modeling voltage oscillations of FIB irradiated devices built with either VO$_{2}$ or V$_{2}$O$_{3}$, both of which are materials exhibiting hysteretic first order phase transitions. These devices are typically composed of thin films of VO$_{2}$/V$_{2}$O$_{3}$ with metallic electrodes patterned via photolithography in a two terminal geometry~\cite{Ghazikhanian23}. The MRN is a mesoscopic scale phenomenological model that represents the thin film Mott material as a rectilinear grid of nanosized cells, where each cell contains four resistors; each resistor is connected in series to the corresponding resistor of the neighboring cell. Given that the crystalline grain size of the thin film is on the order of 10-100 nm, we build the MRN as a 100$\times$106 sized grid of cells. This ensures that each cell of our model corresponds to a mesoscopic portion of the physical device that has a well-defined resistance value $R$. In addition, we include two electrodes within the network that are 42$\times$12 cells large and are ideally metallic. Apart from the electrodes, we assume that every individual cell has the possibility of being either in an insulating or metallic phase. For each resistor, the probability to be in either the high resistance phase $R_{\rm h}$ (insulating) or the low resistance phase $R_{\rm l}$ (metallic) depends on the local temperature of the cell in the network through a Landau free energy functional~\cite{Rocco22a}. Every resistor obeys Ohm's law $V = IR$. When a voltage is applied and current begins to flow across the network, heat is locally generated in each cell according to Joule's first law $P = IV$. Every cell is assumed to be in thermal contact with its four neighbors, and additionally with an ideally insulating substrate that is held to a fixed temperature $T_0$. Heat dissipation may occur through the neighboring cells and the substrate according to the thermal conductivity $\kappa$. The evolution of the temperature of the cell at location $\bm{r} = n\hat{x} + m\hat{y}$ such that $n,m\in\mathbb{Z}$ is then given by
\begin{equation}
    \dfrac{dT_{\bm{r}}}{dt} = \dfrac{P_{\bm{r}}}{C} - \dfrac{\kappa}{C}\bigg[(T_{\bm{r}} - T_0) + \sum_{\bm{d}=\pm\hat{x},\pm\hat{y}} (T_{\bm{r}} - T_{\bm{r} + \bm{d}}) \bigg].
    \label{tempurature}
\end{equation}
Here $C$ is the heat capacity of the thin film.

The model is then solved iteratively as follows: At first, a voltage difference $V_{\rm M}$ is applied across the entire resistor grid. Initially all of the resistors are in the high resistance state $R_{\rm h}$ at temperature $T_0$, and the current through each resistor is calculated via Kirchhoff's laws. These currents define the power $P = IV$ within each cell. Next, the temperature of every cell is updated according to Eq.~\eqref{tempurature}. Given these new temperatures, the resistance value of every resistor has a probability to drop down to $R_{\rm l}$ (or to return to $R_{\rm h}$ from $R_{\rm l}$) according to the transition rate
\begin{equation}
    \nu(T) = \nu_0 \exp[-\Delta_{E}(T)/T],
\end{equation}
where $T$ is the local temperature of the cell, $\nu_0$ is the attempt rate, and $\Delta_{E}(T)$ is either the insulating or metallic energy barrier of the Landau free energy. With these new resistance values the currents across each resistor are once again calculated according to the applied voltage $V_{\rm M}$, restarting the iterative cycle. 

\begin{figure}
    \centering
    \includegraphics[width=1.0\columnwidth]{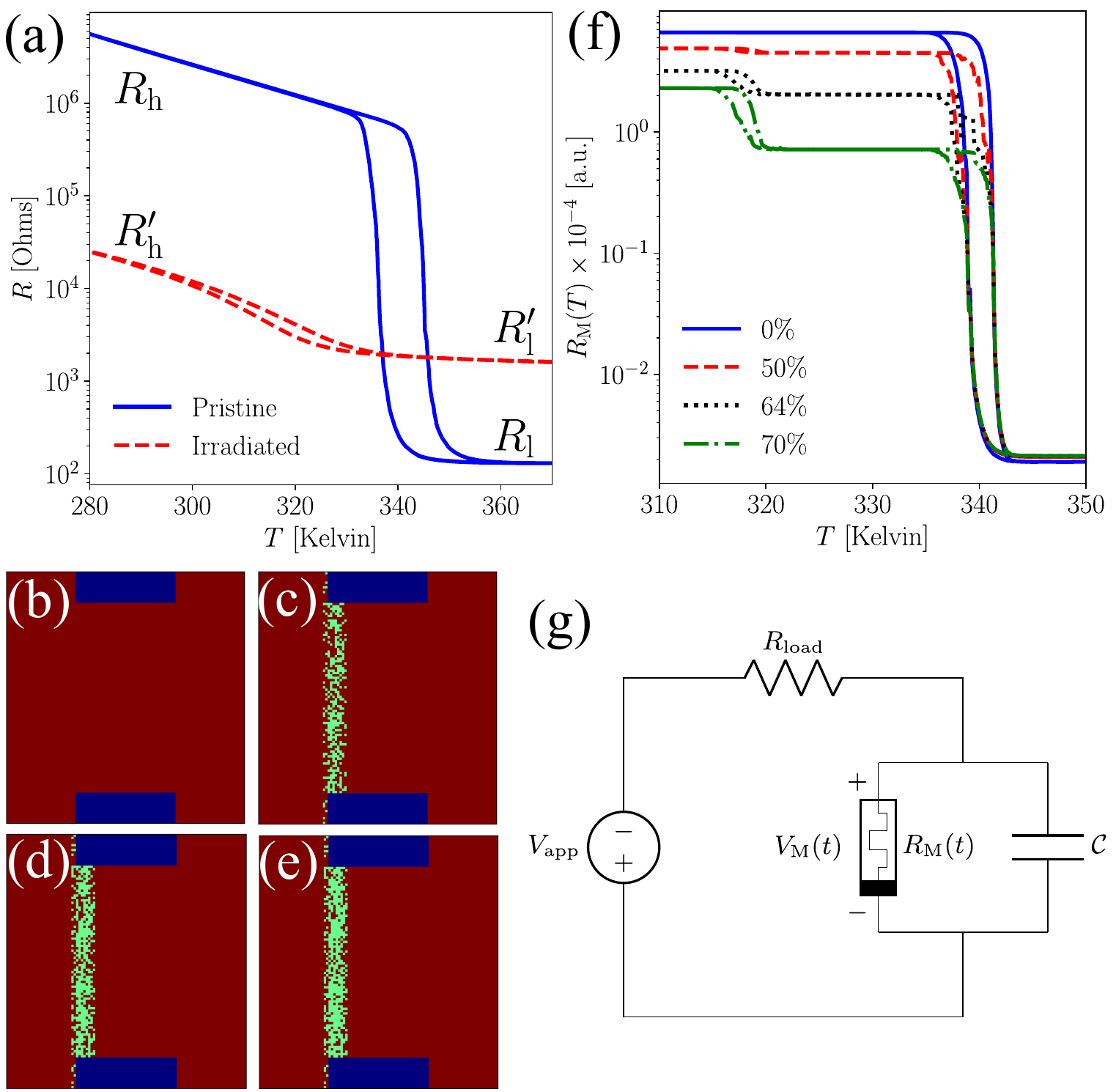}
    \caption{Effects of FIB irradiation on VO$_{2}$. (a) Experimental data showing the resistance versus the temperature of both pristine and blanket irradiated VO$_{2}$, such that the irradiation covers the entire sample, clearly exhibiting a hysteretic first order IMT. (b)-(e) Schematics of the MRN, showcasing the defects (green) induced by FIB irradiation and the electrodes (blue). The defects are randomly placed within a stripe with a density of (b) 0\% (c) 50\% (d) 64\% (e) 70\%. (f) The total resistance of the resistor network $R_{\rm M}(T)$ shown above as a function of temperature, highlighting the first-order transition of the normal cells at 340 K, and also the defect cells at 318 K. The exact parameter values for the defects are given in Sec.~\ref{sec3a}. (g) Schematic of the Mott spiking neuron circuit. A voltage source with a constant applied voltage $V_{\rm app}$ is in series with a load resistor $R_{\rm load}$ along with a parallel set including a capacitor with capacitance $\mathcal{C}$ and the MRN, which has the time dependent resistance $R_{\rm M}(t)$. The voltage drop across the MRN and the capacitor is given by $V_{\rm M}(t)$.}
    \label{blanket}
\end{figure}

% \begin{figure}
%     \centering

%     \begin{circuitikz}[american voltages]
% \draw
%   (0,0) to [V, l^=$V_{\rm app}$] (0,4)
%   to [R,l^=$R_{\rm load}$] (4,4)
%   to [short] (4,3)
%   (3,3) to [short] (5,3)
%   (3,3) to [memristor, l^=$R_{\rm M}(t)$, v=$V_{\rm M}(t)$] (3,1)
%   (5,3) to [C, l^=$\mathcal{C}$] (5,1)
%   (3,1) to [short] (5,1)
%   (4,1) to [short] (4,0) to [short] (0,0)
%   ;
  
% \end{circuitikz}
    
%     \caption{Schematic of the Pearson-Anson oscillating circuit. A voltage source with a constant applied voltage $V_{\rm app}$ is in series with a load resistor $R_{\rm load}$ along with a parallel set including a capacitor with capacitance $\mathcal{C}$ and the MRN, which has the time dependent resistance $R_{\rm M}(t)$. The voltage drop across the MRN and the capacitor is given by $V_{\rm M}(t)$.}
%     \label{tikz}
% \end{figure}

In the above process, the transition rate $\nu(T)$ from $R_{\rm h}$ (insulator) to $R_{\rm l}$ (metallic) smoothly increases from $\nu(T\ll T_{\rm IMT}) \approx 0$ to $\nu(T\gg T_{\rm IMT})\approx 1$ about the material's transition temperature $T_{\rm IMT}$. In Fig.~\ref{blanket}(a), we plot the experimentally measured resistance against the temperature for samples of both pristine VO$_{2}$ and one that has been subjected to blanket Ga$^{+}$ ion beam irradiation. For pristine VO$_{2}$, the transition temperature is given by $T_{\rm IMT} = 340$~K. We may observe that FIB irradiation results in two main consequences: First, the insulating and metallic resistances of the irradiated sample are modified. $R_{\rm h}$ of the irradiated sample decreases by several orders of magnitude to $R_{\rm h}'$, while $R_{\rm l}$ increases by an order of magnitude to $R_{\rm l}'$. Second, the transition temperature $T_{\rm IMT}'$ of the irradiated sample is substantially reduced, such that $T_{\rm IMT}'\approx 318$ K. For even higher levels of irradiation, $R_{\rm h}'$ and $R_{\rm l}'$ continue to approach one another until the IMT disappears.

To account for the consequences of FIB irradiation described above, we may modify the MRN according to Figs.~\ref{blanket}(b) through (e). For the experimental devices shown within this work, FIB irradiation is used to draw filamentary paths in the shape of stripes about $\sim1$ $\mu$m in thickness. Here, cells in green represent ``defects" resulting from FIB irradiation which have two main differences compared to the normal cells in red: First, the green cells may only transition between the resistance values $R_{\rm h}'$ and $R_{\rm l}'$, such that $R_{\rm h} \gg R_{\rm h}' \gg R_{\rm l}' \gg R_{\rm l}$. Second, the transition temperature of the green cells is lowered such that $T_{\rm IMT}' < T_{\rm IMT}$. These modifications of the MRN are consistent with the consequences of Ga$^{+}$ ion beam irradiation as shown in Fig.~\ref{blanket}(a), and with previous studies of FIB irradiation on VO$_{2}$ and V$_{2}$O$_{3}$~\cite{Trastoy20,Ramirez15,Ghazikhanian23}. In Fig.~\ref{blanket}(f), we plot the resistance versus the temperature of the entire MRN device (featuring no capacitor) for various densities of defects within a 10 cell wide stripe pattern, such that the defect cells are randomly placed. It can be seen that significant deviations from ideal behavior occur for defect densities above 50\%, allowing for the possibility to control the location of the metallic filament and the RS behavior within these FIB irradiated regions of the sample. The IMT of the defect cells can be seen in the higher densities at $T_{\rm IMT}'=318$ K, indicating the existence of a ``defect filament" which, while not as conductive as the typical metallic filament, requires less input power to produce and possesses the same volatile RS behavior as the pristine sample.

Although developed here for VO$_{2}$ and V$_{2}$O$_{3}$, we note that this defect-extended MRN framework applies broadly to any volatile RS material, such as the nickelates and other correlated oxides~\cite{Disa13}.

\section{Results}
\label{sec3}

\begin{figure}
    \centering
    \includegraphics[width=\columnwidth]{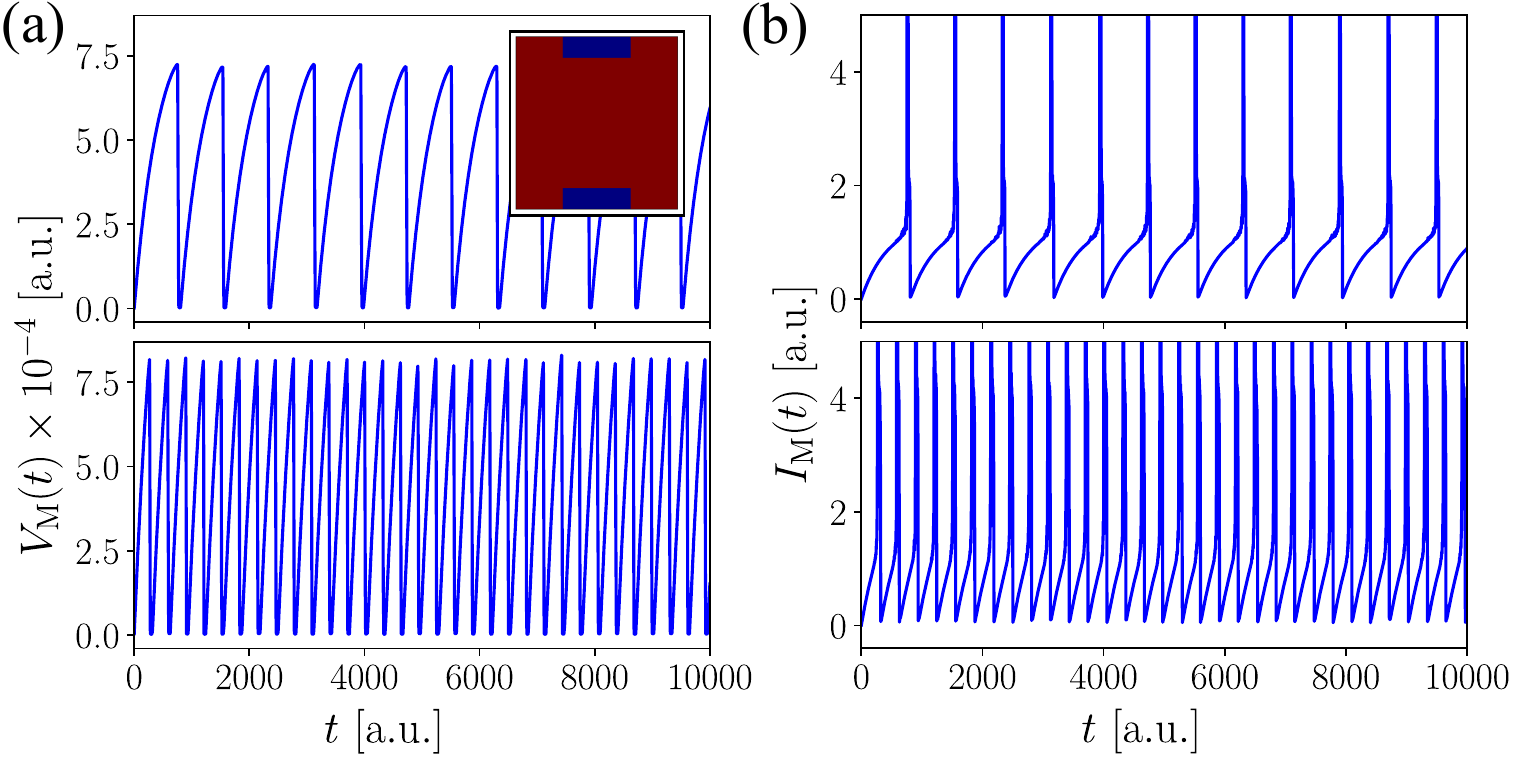}
    \caption{Oscillations of the MRN with 0\% defects when implemented in a Mott spiking neuron circuit, showing the (a) voltage $V_{\rm M}(t)$ (b) current $I_{\rm M}(t) = V_{\rm M}(t)/R_{\rm M}(t)$. From top to bottom, for both the the voltage and current, the applied voltages are $V_{\rm app}/10^{4} =$ 25 a.u. and 50 a.u. respectively. The inset shows a schematic of the pristine MRN.}
    \label{voltosc1}
\end{figure}

We begin with an analysis of the pristine Mott spiking neuron device in the absence of any defects. The MRN is initialized by setting $R_{\rm h}=3.5\times10^{4}$ a.u. and $R_{\rm l}=10$ a.u. (in the MRN all units are considered arbitrary). The substrate temperature is set to $T_{0}=313$ K. Such devices have previously been studied with the MRN in the analysis of V$_{3}$O$_{5}$, which features a second order IMT~\cite{Adda22}. As shown in the schematic in Fig.~\ref{blanket}(g), a capacitor with capacitance $\mathcal{C}$ is added in parallel to the MRN, which is kept constant throughout the rest of this work. A constant applied voltage $V_{\rm app}$ is input to the parallel MRN and capacitor, with $V_{\rm M}(t)$ being the voltage drop across both the MRN and the capacitor. As the capacitor begins to gain charge, $V_{\rm M}(t)$ begins to increase, and continues until it is large enough to trigger a resistive collapse via the formation of a metallic filament connecting the electrodes of the device, as shown in the top panel of Fig.~\ref{voltosc1}(a) at approximately $t_0\approx1000$ a.u. At this point, the resistance of the resistor network $R_{\rm M}(t)$ suddenly and dramatically drops, which allows the capacitor to discharge through the resistor grid as shown in the current spikes in Fig.~\ref{voltosc1}(b). Before the filament formation, the resistance is high, and the time constant of the increasing voltage was proportional to $\tau\propto R_{\rm M}(t<t_0)\mathcal{C}$. However, after the sharp decrease of the resistance, it is clear that the voltage across the capacitor and MRN drops according to the new time constant $\tau_{2}\propto R_{\rm M}(t>t_0)\mathcal{C}\ll\tau$, which may be seen in the voltage oscillations of Fig.~\ref{voltosc1}(a).

The voltage oscillations of the pristine device only occur for a finite range of $V_{\rm app}\in(V_{\rm app}^{\rm min},V_{\rm app}^{\rm max})$, such that approximately $20 \ {\rm a.u.}\lesssim V_{\rm app}/10^4\lesssim 50 \ {\rm a.u.}$ As addressed in Ref.~\onlinecite{Rocco22}, these boundaries are both experimentally and theoretically difficult to define precisely, owing to the inherently stochastic nature of the first order phase transition and of the filament formation and reabsorption. The frequency of the voltage oscillations can clearly be seen to increase with increasing $V_{\rm app}$, such that the average peak-to-peak distance of $V_{\rm M}(t)$ steadily and monotonically decreases. 

For voltages close to $V_{\rm app}^{\rm min}\approx20\times 10^4$ a.u., there is a significant variability in the peak-to-peak time period, giving rise to the uncertainty of the frequency $f(V_{\rm app})$. This variability steadily decreases as $V_{\rm app}$ is increased, and is at its lowest at the upper end of the applied voltage window. For voltages above the voltage window such that $V_{\rm app}/10^4\gtrsim50$ a.u., the device only oscillates a finite number of times before a filament is permanently maintained and $V_{\rm M}(t\to\infty)\approx0$ a.u. The system is not deterministic, and running the simulation again may result in a different number of these oscillations. Even higher values of $V_{\rm app}$ decrease this number until a permanent filament is created within the first voltage cycle. This same behavior is also seen in the depolarization block of biological neurons, which oscillate only a finite number of times when under large injection current~\cite{Bianchi2012}.

\subsection{Stochastic filament lifetimes within V$_{2}$O$_{3}$}
\label{sec3a}

We may then analyze the consequences of adding FIB irradiation-induced defects to the MRN in the arrangements shown in Figs.~\ref{blanket}(c) through (e). Here we set $R_{\rm h}'=500$ a.u., $R_{\rm l}'=100$ a.u., and $T_{\rm IMT}' = 318$ K. At first, for low defect densities, the effects are minimal. From 0\% to 50\% defects within the 10 cell stripe pattern, there is an increase in the window of applied voltages that admit oscillations. At 50\% defects, oscillations occur as $8 \ {\rm a.u.}\lesssim V_{\rm app}/10^4 \lesssim73$ a.u. In addition, the amplitude of $V_{\rm M}(t)$ reduces by a factor of 2. However, at these densities, the frequency $f(V_{\rm app})$ still monotonically increases with $V_{\rm app}$.

For even larger densities, more pronounced differences begin to emerge. At approximately $\sim60\%$, the upper boundary of the applied voltage window begins to drop down suddenly towards $V_{\rm app}^{\rm min}$. We emphasize that the percentage in which this happens depends on the exact \textit{pattern} of defects in the stripe, and different random arrangements lead to different necessary percentages. We comment that this is similarly observed in experiment, where different samples with the same level of irradiation can exhibit different types of oscillation behavior.

\begin{figure}
    \centering
    \includegraphics[width=\columnwidth]{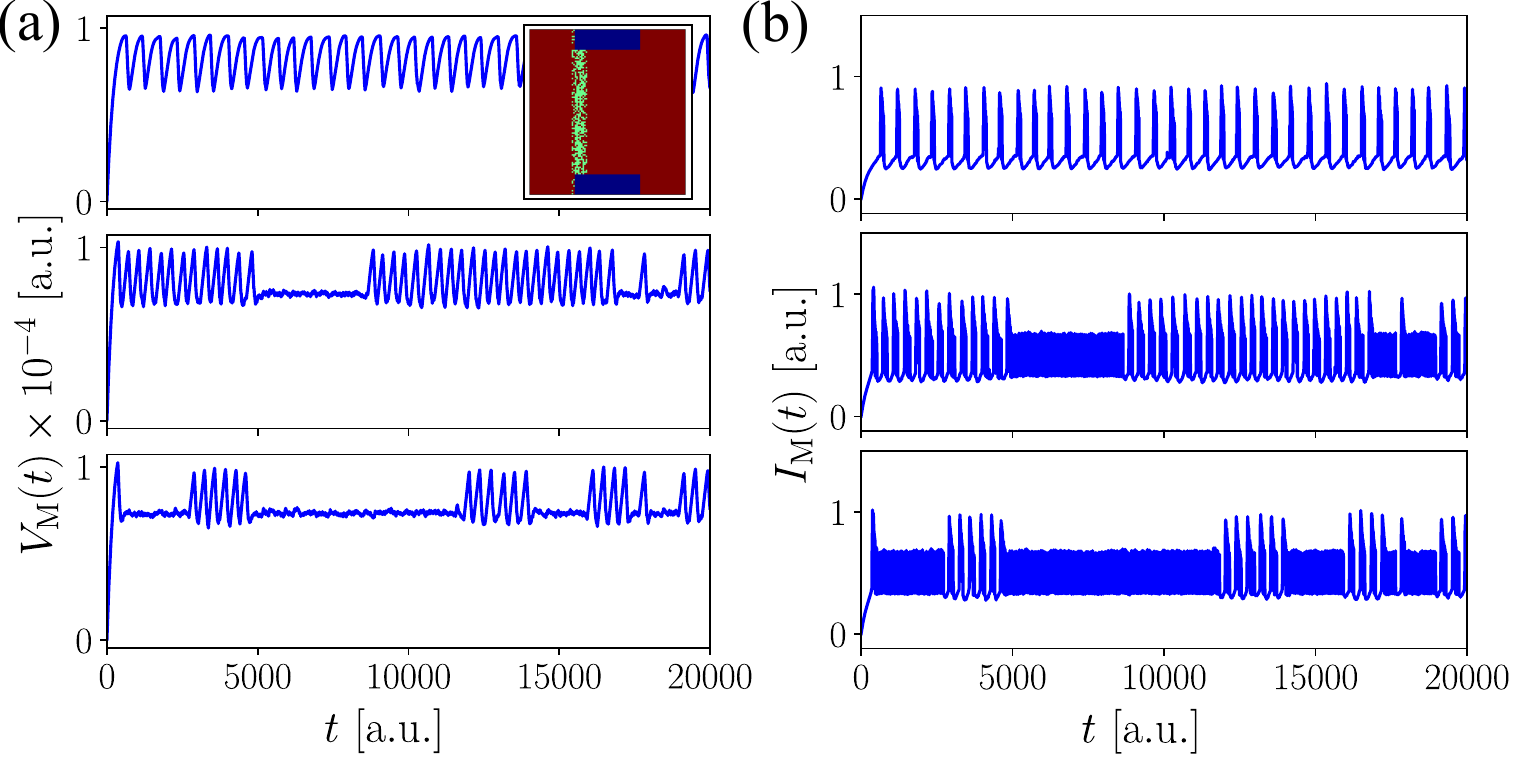}
    \caption{Oscillations of the MRN with 64\% defects defects when implemented in a Mott spiking neuron circuit, showing the (a) voltage $V_{\rm M}(t)$ (b) current $I_{\rm M}(t) = V_{\rm M}(t)/R_{\rm M}(t)$. From top to bottom, for both the the voltage and current, the applied voltages are $V_{\rm app}/10^{4} =$ 5.2 a.u., 6.5 a.u., and 6.52 a.u. respectively. The inset shows a schematic of the irradiated MRN.}
    \label{voltosc2}
\end{figure}

In Fig.~\ref{voltosc2} we plot the voltage and current oscillations of an arrangement of defects at a 64\% density. In this case, oscillations only occur as $5 \ {\rm a.u.}\lesssim V_{\rm app}/10^4 \lesssim 6.52 \ {\rm a.u.}$ and feature clear qualitative differences from the pristine case. The first significant change is that $V_{\rm M}(t)$ never drops towards zero, implying that the capacitor does not fully discharge during the voltage cycle. This larger minimum voltage implies that filament reabsorption is ultimately less likely to occur. We may also notice that the amplitude of $V_{\rm M}(t)$ has decreased by an order of magnitude. The next stark difference can easily be seen as the applied voltage increases towards $V_{\rm app}^{\rm max}\approx 6.52\times10^4$ a.u. The peak-to-peak distance of $V_{\rm M}(t)$ suddenly begins to exhibit large gaps between oscillations. 

\begin{figure}
    \centering
    \includegraphics[width=\columnwidth]{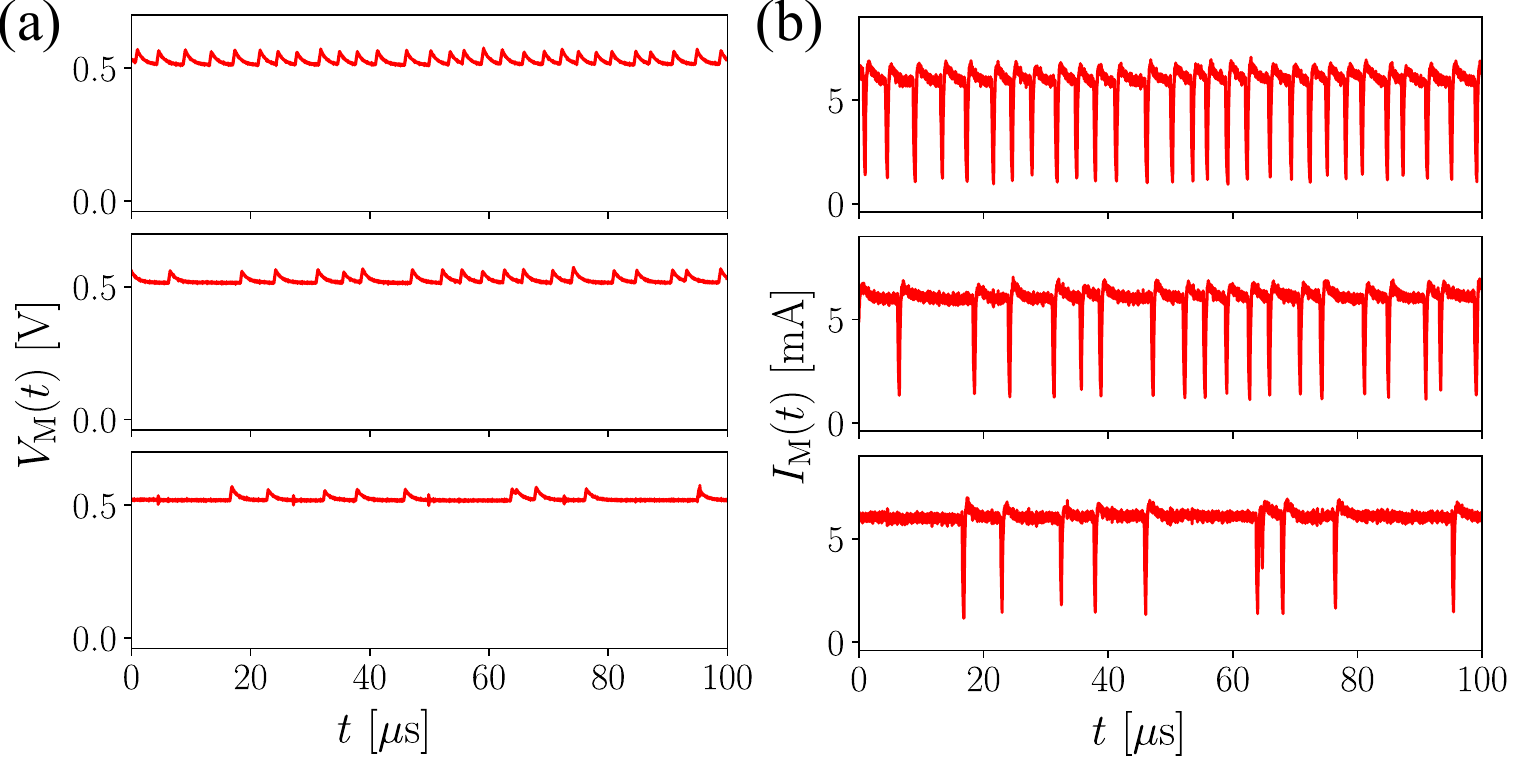}
    \caption{Oscillations of an experimental device featuring FIB irradiated V$_{2}$O$_{3}$ in a configuration similar to Fig.~\ref{blanket}(c), showing the (a) voltage and (b) current. From top to bottom, for both the voltage and the current, the applied voltages to the circuit are $V_{\rm app} =$ 700 mV, 710 mV, and 720 mV respectively.}
    \label{v2o3data}
\end{figure}

These novel features of the MRN featuring defects, such as the stochasticity which increases as the input voltage increases, can be directly observed within experimentally fabricated spiking neuron circuits. In Fig.~\ref{v2o3data} we plot the voltage and current oscillations of a device featuring a V$_{2}$O$_{3}$ sample that has been subject to FIB irradiation in a similar configuration as that of Fig.~\ref{blanket}(c). For pristine V$_{2}$O$_{3}$, the voltage and current oscillations exhibit the same form as what the MRN demonstrates in Fig.~\ref{voltosc1}; periodic oscillations arise from the cyclic formation and reabsorption of a metallic filament, reflected experimentally as regular current spikes whose frequency increases with applied voltage. In the irradiated device, however, periodic oscillations persist at low voltages but evolve into irregular bursts separated by long intervals at higher voltages, indicating increased stochasticity in the peak-to-peak period~\cite{Ghazikhanian25}. Less input power is needed to drive oscillations, the amplitude of the voltage oscillations is lower, and the minimum voltage never drops down to zero, similar to what is predicted by the MRN. The three voltage and current traces of Fig.~\ref{v2o3data} are taken at the upper end of the oscillatory window, just before $V_{\rm app}^{\rm max}$. Just like what is observed in the MRN, as $V_{\rm app}$ approaches $V_{\rm app}^{\rm max}$, the stochasticity begins to rise with larger and more random gaps between individual cycles, characterized by the presence of a persistence and long-lived metallic filament formed along the FIB-induced defects. The close correspondence between simulation and experiment confirms that the MRN model successfully captures the essential behavior of both pristine and defect-engineered volatile resistive switching materials.

\begin{figure}
    \centering
    \includegraphics[width=\columnwidth]{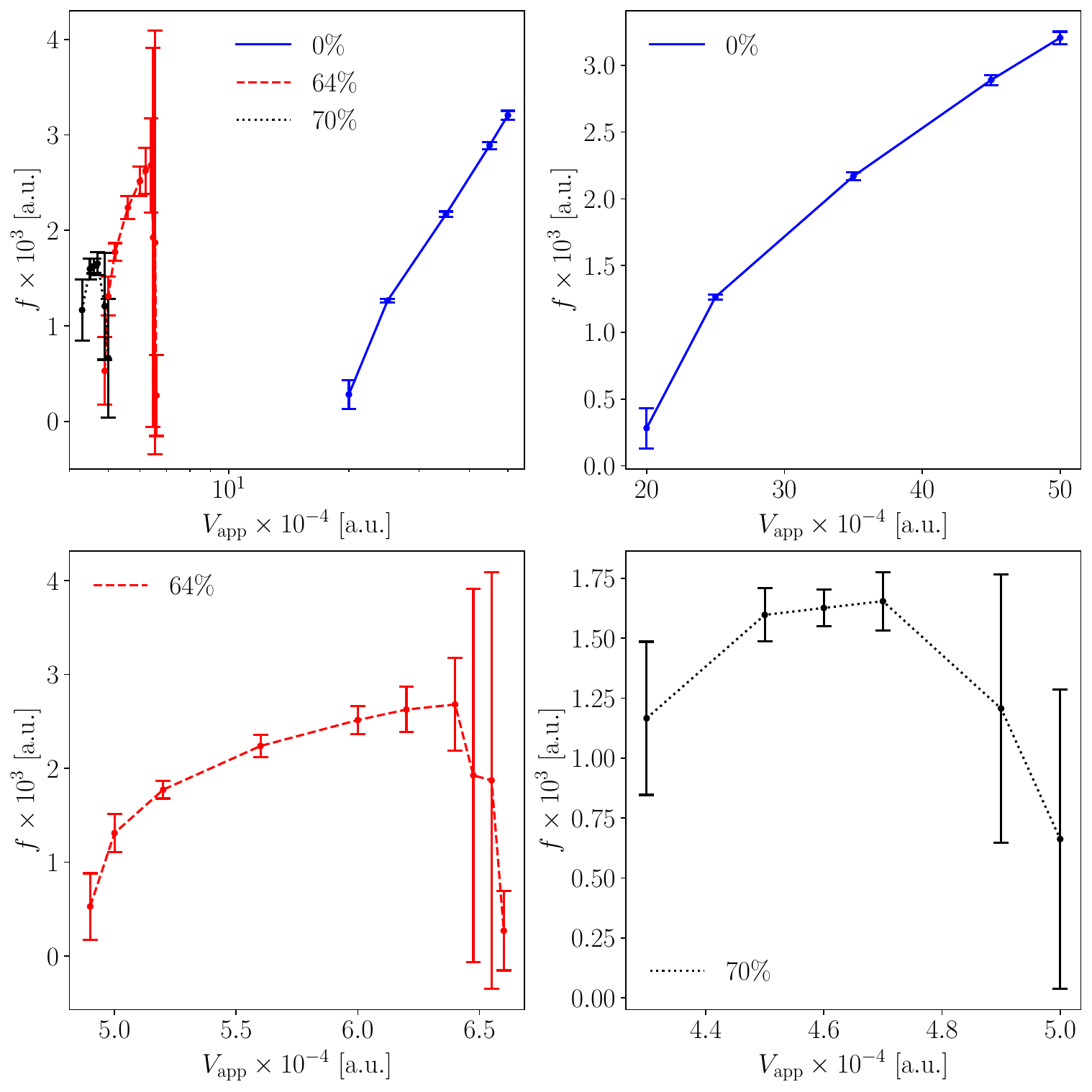}
    \caption{Frequency of the voltage oscillations as a function of the applied voltage, with defect densities given by 0\%, 64\%, 70\%. In the top left panel, the data for all three cases are plotted together on a log scale, while the remaining panels focus on each individual irradiation level.}
    \label{freq}
\end{figure}

These long gaps between brief periods of spiking oscillations arise due to the appearance of a semi-stable yet fragile filament created by the defect cells, which only survives for a finite amount of time before ultimately being reabsorbed. The gaps become both more frequent and longer lasting as $V_{\rm app}$ increases further towards $V_{\rm app}^{\rm max}$. The existence of these semi-stable filaments implies that, in contrast to the pristine case, the cycle-to-cycle variability and stochasticity of the spiking neuron device \textit{increases} with increasing applied voltage, and the average frequency $f(V_{\rm app})$ is no longer monotonic with $V_{\rm app}$. This can be seen in Fig.~\ref{freq}, where not only does the uncertainty of $f(V_{\rm app})$ increase with the applied voltage, but the average value of $f(V_{\rm app})$ itself reaches a minimum at the highest applied voltage. The overall profile of $f(V_{\rm app})$ in the defect-extended MRN framework is qualitatively similar to the behavior of biological neurons under high injection current, as can be seen in the study of the depolarization block of mice neurons in Ref.~\onlinecite{Bianchi2012}.

We also note that in Fig.~\ref{freq}, the frequency at low applied voltages for all levels of irradiation is well-modeled by the leaky integrate-and-fire neuron model $f(V_{\rm app})\propto -[\log(1 - V_{\rm app}^{\rm min}/V_{\rm app})]^{-1}$. Our work therefore demonstrates that FIB irradiation can help further model the high voltage behavior of biological neurons in neuromorphic circuitry.

\begin{figure}
    \centering
    \includegraphics[width=\columnwidth]{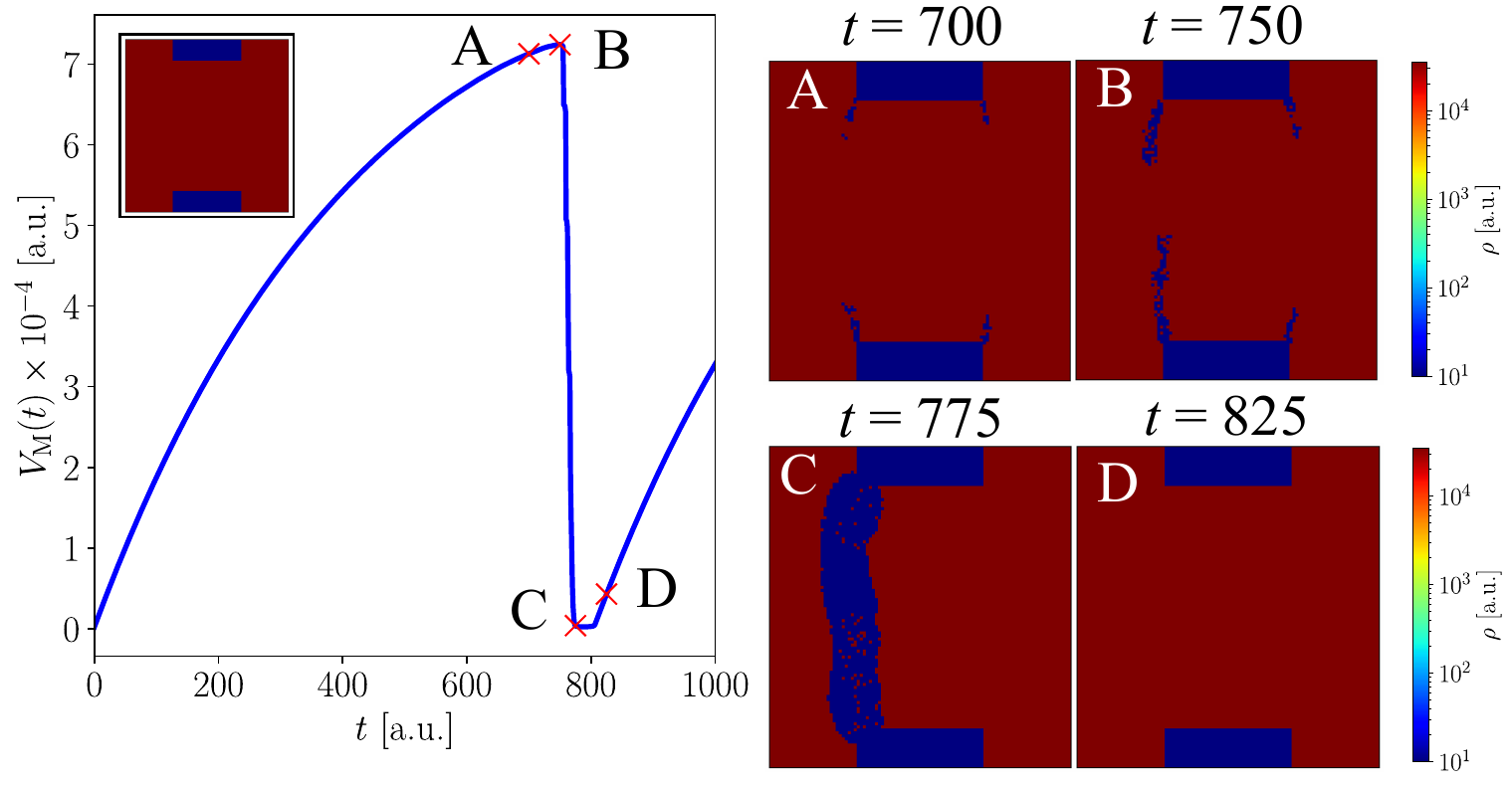}
    \caption{Evolution of the metallic filament within a pristine sample. The left panel plots the voltage of the MRN of the first oscillation given in the top panel of Fig.~\ref{voltosc1}. The four panels on the right show snapshots of the MRN at different times during the cycle. Cells in red are insulating with resistance $R_{\rm h}$, while cells in dark blue are metallic with resistance $R_{\rm l}$. (The rectangular electrodes however are ideally metallic).}
    \label{fil1}
\end{figure}

\begin{figure}
    \centering
    \includegraphics[width=\columnwidth]{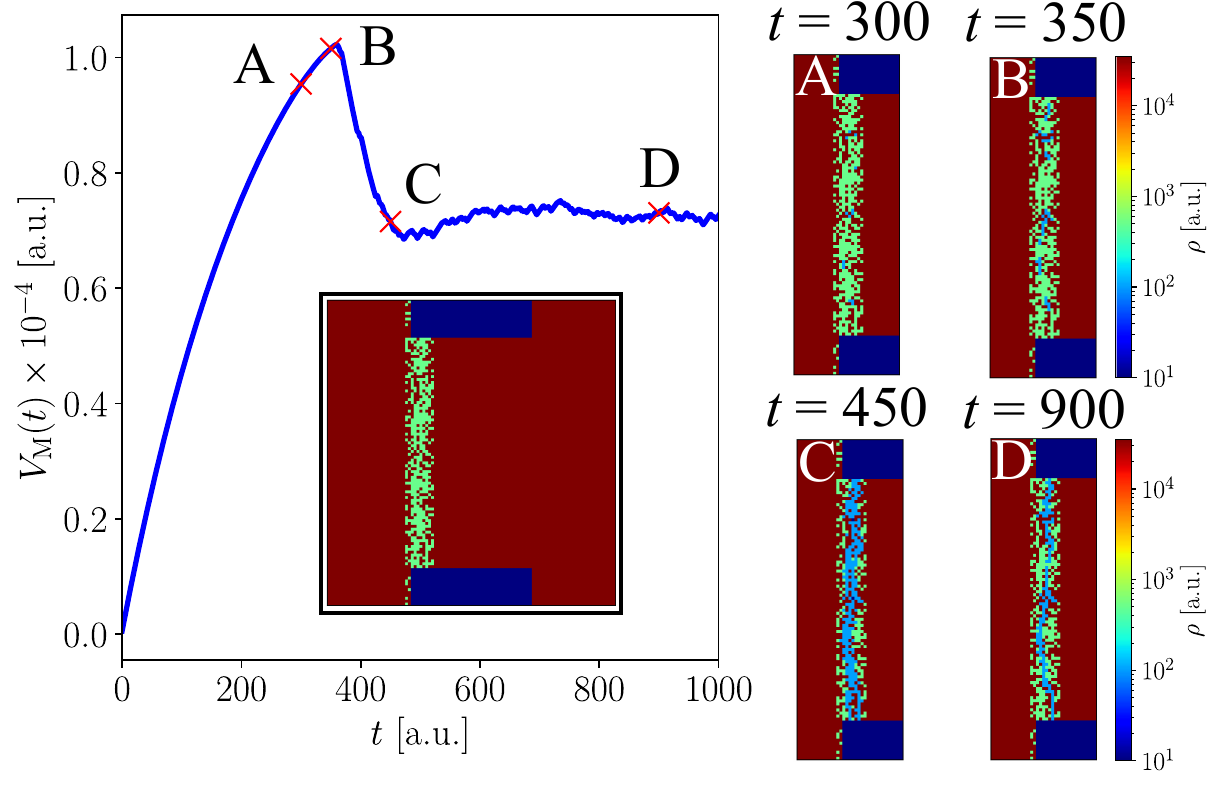}
    \caption{Evolution of the metallic filament within a sample features FIB irradiation-induced defects. The left panel plots the voltage of the MRN of the first oscillation given in the bottom panel of Fig.~\ref{voltosc2}. The four panels on the right show snapshots of the MRN at different times during the cycle. Cells in green are insulating defects with resistance $R_{\rm h}'$, while cells in light blue are metallic defects with resistance $R_{\rm l}'$.}
    \label{fil2}
\end{figure}

In Figs.~\ref{fil1} and \ref{fil2} we plot snapshots of the filament formation processes for the pristine and irradiated cases respectively. In Fig.~\ref{fil1} we plot the first voltage oscillation of the data given in the top panel of Fig.~\ref{voltosc1}(a) showing both the filament formation and reabsorption, while in Fig.~\ref{fil2} we plot the first voltage oscillation of the data given in the bottom panel of Fig.~\ref{voltosc2}(a) showing the long lasting survival of a filament.

In Fig.~\ref{fil1} the filament growth is shown to be gradual and predictable. Metallic cells form around the corners of the electrodes as a consequence of the increased electric field density due to the edge effect. These metallic cells form slow growing filamentary structures that eventually meet in the middle of the device, at which point the current spikes, the metallic filament grows and covers a large portion of the device, and the voltage $V_{\rm M}(t)$ drops dramatically. As detailed in Ref.~\onlinecite{Rocco22}, this slow and deterministic filament growth at the lower end of the $V_{\rm app}$ oscillation window is a consequence of the value of the thermal conductivity $\kappa$ (or, equivalently, the ratio between $R_{\rm h}$ and $R_{\rm l})$. For larger values of $\kappa$, the filament growth process at low $V_{\rm app}$ will exhibit more stochasticity and an avalanche-like behavior. After the voltage $V_{\rm M}(t)$ drops towards zero, the filament is then quickly reabsorbed.

In Fig.~\ref{fil2} we see qualitatively different behavior. In this case, the metallic filament is confined within the irradiated zone and principally forms along the FIB irradiation-induced defects which drop to the resistance value $R_{\rm l}'$, as shown in light blue. The filament formation also begins within the middle of the sample, far from the electrodes and in between the gaps of adjacent defect cells. To complete the filament, non-defect cells with resistance value $R_{\rm l}$ (shown in dark blue) fill the space between nearby defects. After the filament is formed and the voltage drops, the metallic filament quickly diminishes in size. However, unlike in the pristine case, here there are enough defect cells to help the filament survive despite $V_{\rm M}(t)$ being reduced. This thin filament eventually becomes reabsorbed after a lifetime significantly larger than the oscillation period, as shown in Fig.~\ref{voltosc2}. The statistics of the lifetimes of these fragile filaments are strongly non-gaussian, and contribute to the growing stochasticity of the spiking neuron device with increasing $V_{\rm app}$. As the minimum of the voltage $V_{\rm M}(t)$ across the network increases, the percentage of filaments with larger than typical lifetimes increases dramatically. These devices, with their lower required input power along with their increased stochasticity, are therefore well-suited for biological modeling and applications in neuromorphic circuitry.

\subsection{Stochastic early firing within VO$_{2}$}
\label{sec3b}

The long-living and stochastic filamentary behavior reported in the previous section is just one of many possible novel behaviors that can emerge from FIB irradiation within volatile RS Mott-based devices. Particularly, this behavior emerges due to the sufficient separation of $R_{\rm h}'$ and $R_{\rm l}'$, allowing for the stabilization of the long-lived filaments mediated via the defect cells. Furthermore, from Fig.~\ref{fil2}, it is clear that the voltage oscillations of the irradiated device (both the periodic and stochastic) arise principally from the transitions of the defect cells. That is to say, in the irradiated case, the filament remains confined within the irradiated region of the sample. We note that if we simply modeled the irradiated defects such that $R_{\rm h}' = R_{\rm l}'$’, the stochastic oscillating behavior would not occur. In this case, the added defects would only decrease the overall resistance of the device and act to localize the location of the filament. On the other hand, to model systems that are less sensitive to the effects of FIB-irradiation, we can increase the separation of $R_{\rm h}'$ and $R_{\rm l}'$. In this case, if $R_{\rm l}'$ is too low, the capacitor will discharge too quickly, allowing the device to cool and reabsorb the defect-mediated filament, preventing it from being long-living and exhibiting the stochastic behavior shown in the previous section.

\begin{figure}
    \centering
    \includegraphics[width=\columnwidth]{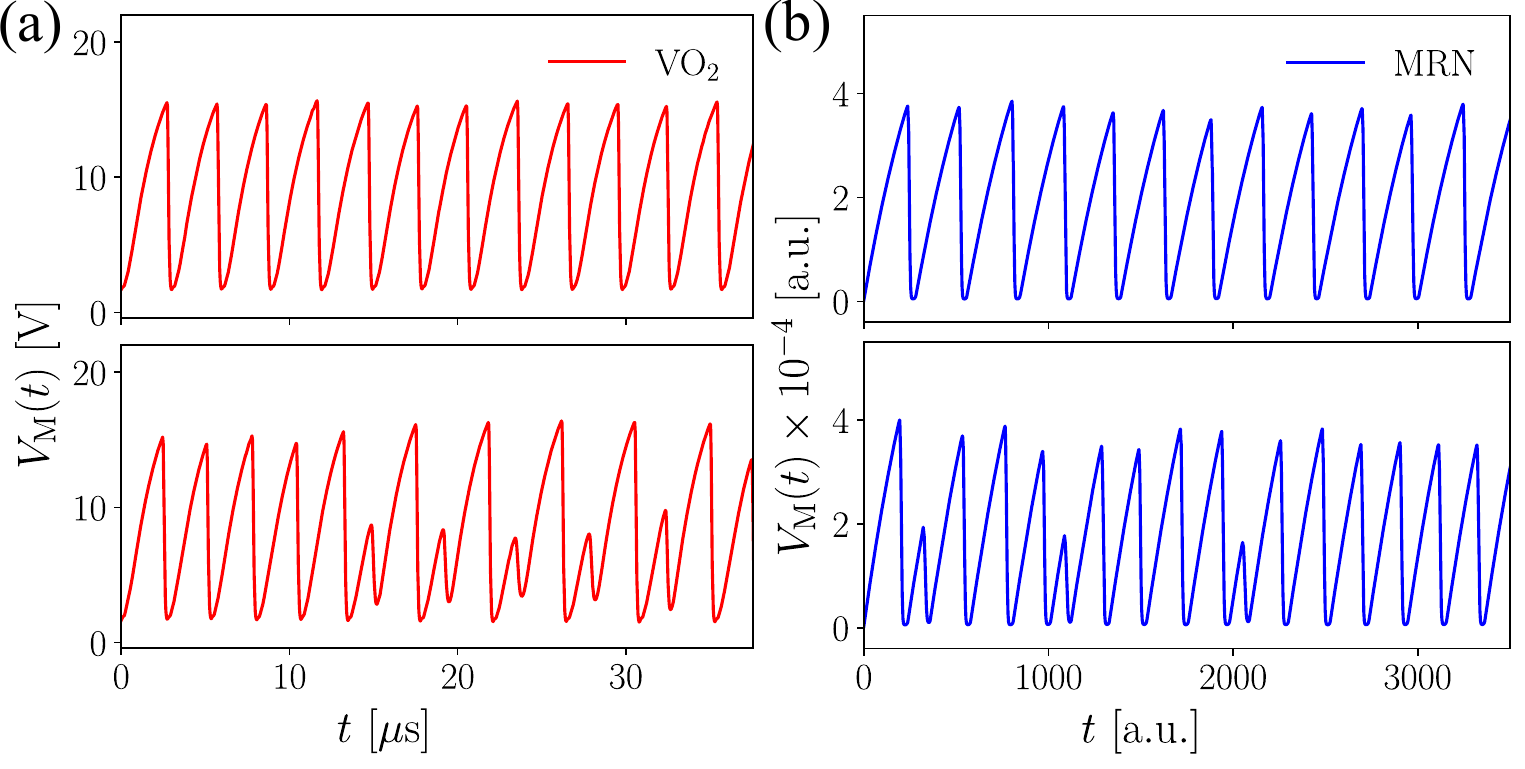}
    \caption{Voltage oscillations of (a) an experimental device featuring FIB irradiated VO$_{2}$ in a configuration similar to Fig.~\ref{blanket}(c), and (b) the MRN. In (a), from top to bottom, the applied voltages are 900 mV and 950 mV respectively, while for (b) the applied voltage to the MRN is $V_{\rm app}/10^{4}=25$ a.u. and 31 a.u. respectively.}
    \label{vo2data}
\end{figure}

The IMT of VO$_{2}$ has been shown to be much less sensitive to the presence of localized disorder and FIB irradiation compared to that of V$_{2}$O$_{3}$, whose resistivities and transition temperature can be modified quite drastically~\cite{Ghazikhanian23}. In Fig.~\ref{vo2data}(a) we show experimental data of the voltage oscillations of a similar spiking neuron circuit as in the previous section, except now featuring a FIB irradiated sample of VO$_{2}$. As before, in the pristine case VO$_{2}$ exhibits periodic oscillations similar to what is shown in Fig.~\ref{voltosc1}(a). However, once irradiated, the circuit begins to display stochastic behavior markedly distinct from that of V$_{2}$O$_{3}$. While the oscillations at low input voltages close to $V_{\rm app}^{\rm min}$ behave similarly to the pristine case, the oscillations at higher applied voltages show novel behavior. The applied voltage of the traces in Fig.~\ref{vo2data}(a) are both in the upper end of the oscillatory regime, just before the threshold voltage $V_{\rm app}^{\rm max}$. Here we can see that as the applied voltage increases, shorter and irregularly occurring ``mini"-oscillations begin to appear. There are no instances of long-lived metallic filaments with stochastic lifetimes, as in the previous section. Instead, at every voltage cycle, there is a small probability that the metallic filament formation may happen unusually quickly. As before, the amplitude of these oscillations are smaller than the pristine case as well. This shows that the presence of localized disorder arising from FIB irradiation can result in a variety of physical phenomena, opening the possibility for a diverse range of potential functional behaviors.

To capture this behavior, we can adapt the MRN to account for the reduced sensitivity of VO$_{2}$ to the presence of localized disorder. In Fig.~\ref{vo2data}(b), we set $R_{\rm h}'=2.5\times10^4$ a.u., $R_{\rm l}'=20$ a.u., $T_{\rm IMT}'=305$ K, $T_{0}=300$ K, and we set the density of the column of defects to 66\%. As in the experiment, the applied voltages of the traces shown in Fig.~\ref{vo2data} are at the upper end of the oscillatory regime. In this case, the defects allow for a new regime of increased stochasticity as the input voltage is increased. Similar to what is seen in the experiment, the presence of the defects allows for the sudden and comparatively quicker formation of a metallic filament after a random number of voltage cycles. The voltage amplitude is not constant from cycle to cycle and the minimum voltage of each cycle drops down to zero, unlike what is seen in V$_{2}$O$_{3}$ and implying that the metallic filament is fully reabsorbed after every voltage cycle. This shows that the stochasticity arising from FIB irradiation, depending on the material, can result in a range of behaviors with many possible applications.

\section{Discussion}
\label{sec4}

In this work we study the consequences of FIB irradiation-induced defects on self-oscillating Mott spiking neuron devices that feature the materials VO$_{2}$ and V$_{2}$O$_{3}$. We have shown that defect engineering via the introduction of controlled disorder by means of FIB irradiation can greatly reduce the switching power required for metallic filament formation within these neuromorphic devices. Furthermore, we numerically and experimentally demonstrate that these defects can introduce largely stochastic and nonperiodic spiking behaviors that have potential uses for both the biological modeling of neurons and for the hardware-level implementation of neuromorphic circuitry. By adjusting the resistivity and transition temperature of randomly placed defects, we have extended the phenomenological Mott resistor network model to simulate the RS behavior of FIB irradiated self-oscillating devices built from both VO$_{2}$ and V$_{2}$O$_{3}$. The pristine devices exhibit oscillations with a frequency that monotonically increases with the applied dc bias voltage. For the simulations modeling FIB irradiated samples of V$_{2}$O$_{3}$, this is not the case, but instead the frequency of oscillations $f(V_{\rm app})$ reaches a maximum for some applied voltage within the allowable window $V_{\rm app}\in(V_{\rm app}^{\rm min},V_{\rm app}^{\rm max})$. The uncertainty of this frequency also sharply increases as $V_{\rm app}\to V_{\rm app}^{\rm max}$ due to a strongly stochastic oscillating behavior that may have important applications in neuromorphic circuitry. This is similar to what is observed in samples of biological neurons, such as those harvested from rats in Ref.~\onlinecite{Bianchi2012}. In this context, the behavior of the upper end of the applied voltage window is referred to as the “depolarization block,” i.e., when the spiking neuron stops firing as the applied voltage is too large. Within the biological samples, the behavior of the depolarization block is not at all uniform. Some neurons quickly stop firing after a threshold voltage is reached, while others exhibit a slower decay of their firing rate. This underscores the notion that a diverse range of behaviors is a central feature in the function of the brain. Therefore, in the case of artificial neurons, a similarly diverse range of spiking behavior can have far reaching consequences for potential neuromorphic functionalities.

We also observe that the FIB irradiation-induced defects can encourage metallic filament formation far from the electrodes within the middle of the sample, and provide a semi-stable environment for long-lasting filaments to survive in, between brief periods of oscillations. For VO$_{2}$, we have shown both experimentally and numerically via the MRN that FIB irradiation can give rise to a qualitatively different type of stochasticity, where the metallic filament formation can occur abnormally quickly after a random number of voltage cycles. For both materials, we find that the irradiated Mott spiking neuron devices exhibit oscillations via their own parasitic capacitance, indicating a potential path towards the miniaturization of Mott neuromorphic circuitry. Our work therefore opens new possibilities in the use of volatile Mott RS materials in the construction of biologically relevant low-energy neuromorphic devices. More generally, this work reveals how local disorder reshapes nonequilibrium dynamics in volatile RS materials, providing a new route to stochastic behavior in correlated oxides and neuromorphic hardware.

\acknowledgments
This research was  supported by the Quantum Materials for Energy Efficient Neuromorphic Computing (Q-MEEN-C), an Energy Frontier Research Center funded by DOE, Office of Science, BES under Award \#DE-SC0019273. Research at the San Diego Nanotechnology Infrastructure (SDNI) of UCSD, a member of the National Nanotechnology Coordinated Infrastructure, is supported by the National Science Foundation (Grant ECCS-2025752). We thank P. Salev for useful discussions.

\bibliography{bibfile}

\end{document}